# RECOMMENDATION OF THE VAO SCIENCE COUNCIL (VAO-SC) FOLLOWING THE MEETING OF JULY 27-28, 2011

THE VAO-SC: G. FABBIANO (*SAO*, CHAIR), C. BROGAN (*NRAO*), D. CALZETTI (*U. MASS AMHERST*), G. DJORGOVSKI (*CALTECH*), P. ESKRIDGE (*MINNESOTA STATE U.*), Z. IVEZIC (*U. WASHINGTON*), E. FEIGELSON (*PENN STATE U.*), A. GOODMAN (*HARVARD*), B. MADORE (*CARNEGIE*), M. POSTMAN (*STScI*), A. SODERBERG (*HARVARD*), T. RECTOR (*U. OF ALASKA*)

## OVERALL CONCLUSIONS

The VAO (Virtual Astronomical Observatory) Science Council (VAO-SC) met on July 27-28, 2011 at the Harvard-Smithsonian Center for Astrophysics in Cambridge MA, to review the VAO performance during its first year of operations. In this meeting the VAO demonstrated the new tools for astronomers that are being released in September 2011 and presented plans for the 2$^{nd}$ year of activities, resulting from studies conducted during the first year.

A year ago (see the 2010 VAO-SC report; Fabbiano et al 2010, arXiv:1006.2168) the VAO-SC had expressed concerns on the approach of the VAO and generated a number of recommendations for the first year program. Similar concerns and closely parallel suggestions were given in the report of the NSF-NASA review committee, chaired by Dr. Pamela Gay, following a meeting held on April 29, 2011 at NSF. Given that the two independent reviews were largely based on pre-release material (i.e., the original VAO plans in the case of the VAO-SC 2010 report, and the Program Execution Plan (PEP) in the case of the Gay report) it is not surprising that they are in agreement. However, a great deal has changed in the interim. The Gay committee did not have the opportunity to review tools and services that are now mature enough to be released this fall. We therefore feel that the positive and current response of the VAO to our Science Council recommendations of 2010 also answers many of the concerns expressed in both our 2010 report and the Gay report.

Based on the results presented by the VAO this July 2011 in Cambridge, the VAO-SC is extremely pleased with the positive response of the VAO, its high performance, timely deliveries after this first year of activity, and vision for the future of interoperable multi-domain data analysis. The VAO has listened to science input and has delivered. The VAO is on the right track to meeting crucial needs of the astronomical community.

We strongly urge the funding agencies to review the impressive results of this year's activities, and continue fully supporting the VAO. The impact of the VAO on the community is clearly within close reach, and support should continue so as not to slow the momentum and have a negative impact on the access to interoperability by the astronomical community.

# COMMENTS ON 1ST YEAR'S VAO ACHIEVEMENTS AND RECOMMENDATIONS

*Members of VAO-SC present at the July 27-28 2011 meeting*: Fabbiano, Brogan, Eskridge, Feigelson, Goodman, Madore, Postman, Soderberg

*VAO 1st year Products and Study Projects*
The VAO team presented demos of the software that is scheduled for release in September, 2011 and the results of several studies addressing future development areas. The demonstrated software included, in the order of presentations: the SED tool IRIS, the Discovery Tool (VAO Portal), Cross-comparison/matching services, Time-Series tools/services, VOSpace, VO-IRAF integration, and TAP Client. The Study Projects included: new developments in ADS to implement content-based searches, the SkyAlert-VO integration; Data-to-Knowledge (Data Mining); and Desktop integration. The VAO Director, Dr. R. Hanisch, tasked the VAO-SC with providing a priority-rated list of recommendations for 2nd year development, to be used for immediate guidance in generating the 2nd year PEP.

The VAO-SC commends the VAO team for the first year's achievements, and for the clarity and range of the presentations. It is clear that the VAO has taken the VAO-SC advice of 2010 at heart and has responded to the 2010 recommendations to focus on software for science analysis rather than further development of database-oriented infrastructure.

We believe that the first year's release will be welcomed by the scientific community, and we recommend a number of these efforts for high priority second-year completion: Tap client; IRIS; Portal; integration of IRAF and other major astronomical software packages into the VO. This last effort will require coordination with the involved Data Centers. The ADS development is also recommended for inclusion in the 2nd year Portal development.

The VAO-SC concluded that the Cross-matching effort has reached a sufficient level of maturity for now. However, the VAO-SC noted that the Cross-matching software presented at this meeting was different from the plans described in the 2010 meeting. While improvements in the future could be foreseen in this area, we recommend that the VAO integrate the existing cross-matching services fully in the upcoming release, and postpone future development discussion to a later time.

The SkyAlert-VO integration study proposed that the VAO develop a Transients Facility, by improving the scalability and functionality of the SkyAlert service with VAO development in this area. The time domain is an exciting growing area in astronomy, however, given the necessity of setting priorities on 2nd year development, the VAO-SC concluded that the full-blown development of a VAO Transient Facility is not of immediate urgency for the VAO. This facility will be important to support LSST users, but the LSST deployment is still a few years ahead. Recognizing SkyAlert, in its present state, as a good and useful service, the VAO-SC recommends that SkyAlert be integrated in the VAO in the 2nd year development with the present level of functionality, so that the VAO user community may be made fully aware of its capabilities. This integration will provide the necessary base for a future VAO Transient Facility development.

The Data-to-Knowledge study presentation discussed a thorough comparison of existing services. As identified in the 2010 VAO-SC report Data-to-Knowledge (or Data Mining) is an important and potentially transformative field for the future of astronomical research, which will be dominated by extensive surveys covering the entire wavelength and time domains. Data-to-Knowledge includes both advanced statistical methods and methods for recognizing patterns and clustering in multi-dimensional data spaces that can offer insight on hidden astrophysical parameters. Being at the focus of data interoperability, the VAO is clearly in the best position for fostering these approaches and introducing them in a timely fashion to the astronomical community. However, based on the status of the study, the VAO-SC concluded that this area is not yet ready for immediate development. We recommend that the team prepare a definite and focused plan, discussing in detail the scientific benefits of the proposed services, and clearly identifying both the inclusion of existing software and the requirements for VAO software development.

These tools and interfaces will integrate VO capabilities in the everyday work of astronomers, provide easy access to data and allow data and tool interoperability on a hitherto inexperienced scale. The 'seamless astronomy environment' that the VAO-SC advocated in 2010 is becoming a reality.

*VAO Science Collaborations*
The VAO has established two science collaborations: CANDELS, based on the eponymous Hubble multi-cycle treasury program, with the CANDELS science team; and the Spectrally Merged Catalog of the Small Magellanic Cloud (SMC**2), aimed at synthesizing multi-wavelength catalogs and images of the SMC, led by Dr. Barry F. Madore of the Carnegie Observatories. Work is proceeding in both areas and it is anticipated that results will be released at the January 2012 AAS meeting. Feedback from these teams is also useful for the development activities. These are in-kind collaborations with no funds exchanged, where the VAO provides services for establishing the VO-compliance of the science products. A less formal science collaboration has been established with Dr. Stan Metchev of SUNY-Stony Brook aimed at a search for brown dwarfs.

The VAO-SC commends the establishment of science collaborations, as a way to both assisting the community in their scientific effort, and to validate the VO. This type of collaboration with 'users' should continue as part of future VAO activities. However, the VAO-SC recommends that for future collaborations, the VAO implement a more transparent solicitation and selection process. It is the understanding of the VAO-SC that the VAO Director, Dr. Hanisch, is fully in agreement with this recommended approach.

Moreover, recognizing the tremendous amount of detailed multi-wavelength as well as time domain data available for M31, the VAO-SC recommends that an M31 special project solicitation for in-kind collaboration be issued by the VAO. The unique merger of results would be of great value to the community.

The VAO Director, Dr. Hanisch, also reported the establishment of a memorandum of understanding with the LSST project. The VAO-SC commends these efforts of engaging the upcoming major astronomical surveys in the VO.

*Promulgating tools and methods for astronomical research through VO compliance of innovative and important software – the 'VAO-Inside' approach*

The VAO-SC feels that the VAO staff should be devoted to code development only when an important capability does not exist elsewhere. The VAO staff should emphasize VO-compliancy of existing software tools, and work with external research groups and data centers to accomplish this end.

The VAO-SC recognizes that the VAO has already been responsive to this concern. A successful example of this type of interaction in the first year of activities is the upgrade of the Penn State VOStat service to VO-compliance; Prof. Feigelson, of Penn State, initiated an unsolicited proposal to the VAO and obtained expert help to successfully upgrade the code to VO standards in a short 2-week project. We hear from Dr. Hanisch that the VAO is exploring a similar collaboration with Drs. Connolly and Krughoff of U. Washington aimed at hosting their ASCOT user environment in the VAO. We call this approach the 'VAO-Inside': including the research community at large in the VAO. Fig. 1 shows the present paradigm of VAO development, where most of the tasks are accomplished by the VAO team (Left), versus the more open paradigm in which the VAO team works in continuing collaborations with expert groups throughout the community (Right).

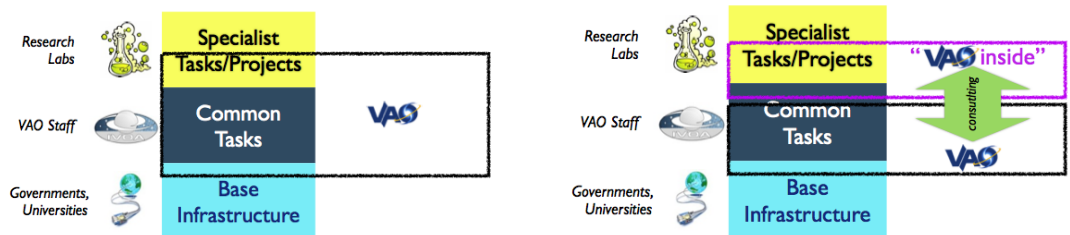

Figure 1. Left: the present VAO paradigm; Right: VAO and VAO-Inside collaborations

We recommend that this type of activity be either actively solicited by the VAO, when an expert group of important software is identified, or managed via a call for proposals, as in the case of science project. The VAO Director has indicated that the

VAO will follow this process during the 2nd year; the VAO-SC may have a role in reviewing these proposals. As in the science collaborations, the VAO should provide mostly in-kind support to established teams. All VO-compliant tools should continue to be advertized and inter-connected, including those generated in-house at the VAO, those by non-US VO groups, and those by external research groups.

Extending the theme of adopting appropriate external software to facilitate science analysis of VAO data products, the VAO-SC encourages the VAO to formally evaluate external packages in data visualization and statistical analysis. Examples include the visualization toolkit VTK and its associated package 3DSlicer developed by NIH for medical imaging, and the R statistical software environment with ~3000 add-on packages. These and other packages are in the public domain and would probably not be fully integrated into VO compliance. If adopted, the astronomical community would inherit extensive software capabilities supported by large external communities of researchers and computer scientists. VAO staff would need only develop interface software such as FITS I/O and IDL AstroLib capabilities, and would then devote additional time to documentation and community training.

*Engaging the professional astronomers*
Professional outreach must be a continuing effort of the VAO and is essential to ensure the ultimate success of this project. We applaud the shift of emphasis in the Project Scientist activities to professional outreach, and we re-iterate the recommendation of the 2010 VAO-SC report in this area. The VAO is a mediating organization in a de-centralized world, providing standards and facilitating collaborations. The scientific community must be made aware of these opportunities and also of the new services developed by the VAO for scientific data analysis and archive access. The VAO-SC recommends that the VAO engage in a series of professional outreach activities: AAS town-hall meetings, workshops for astronomers at geographically dispersed high-leverage institutions, producing demos and distributing them in various media (e.g., videos posted on YouTube), taking care that documentation is clear and easily accessible, and using the VAO web page for disseminating this information as well. *As often as is possible,* VAO experts should be used to introduce astronomers to, and to use, the most modern "online astronomy" tools in a variety of settings—with an emphasis on one-on-one interactions, in-person and also online through a (new) help ticket option at the VAO Portal.

*Engaging the amateur astronomers, the K-12 educational community and the general public*
The VAO-SC recommends that the VAO engage in outreach collaborations, aimed at the general public and at learners of all ages. The wild success of Google Sky and WorldWide Telescope (WWT), which were started independently from the VAO, is a testament to the public's interest in astronomy, and in "real data." WWT, in particular, uses VAO services extensively, making it a powerful tool for professionals and amateurs alike. The rich interconnections to VO tools (via SAMP) and to ADS that WWT offers professionals are paralleled by the equally-rich inter-connections offered to non-professionals, such as direct links to Wikipedia for all objects on the sky and links from within an all-sky environment to image archives hosting high-resolution views. Educational programs like WorldWide Telescope Ambassadors (independent of the VAO) have now demonstrated the power richly-linked research data in classrooms and other learning environments. It is critical that NSF and NASA capitalize further on the VAO investment by further enhancing—and promoting—the VAO's value to STEM education in the United States. The cost/benefit ratio of this investment is very low, and its impact—both on learners and on politicians—will be very high.

*Publicizing the hidden VO*
VO infrastructure is becoming increasingly common in the services provided by established NASA (e.g., HEASARC, CXC, MAST, STScI, IPAC) and NSF funded (e.g., NOAO, NRAO, ALMA, LSST) Data Centers; the VAO collaboration continues to have a central role in facilitating these developments. The journals are moving to be VO-compliant when appropriate (e.g. online tables), and the implications of this option thus far have not been recognized by most astronomers. Thus another opportunity exists for community cooperation in this area as well. But, this enabling VO-compliance is not readily apparent to most astronomers. While the access and interoperability provided is recognized and appreciated, the VAO efforts go largely unrecognized and under-appreciated. This might be seen as good because it demonstrates that the VO access is successful. However, it is important that these contributions and developments be made clear to the funding agencies. (In the words of former SDSC Director Fran Berman, "the best infrastructure is invisible, making its value very hard to 'show'.") The VAO-SC recommends that the VAO survey all major data centers and compile lists of VO infrastructure and services commonly used. This information should then be provided in the VAO web page.

*Establishing a clear process for managing VAO tools and services*
The VAO is currently developing services and tools and is about ready to release them to the community. The VAO sponsored software, once development is completed, will still need to be maintained operationally. This will require some sustained effort on the part of the VAO. Software left untouched for a long time will go stale, at the very least because the operating systems will evolve underneath it. The VAO-SC advocates that for the duration of the VAO any package that is deemed important and is clearly used by the community should be kept alive. Generally maintenance is a minor cost compared to development, assuming that no new features are developed. The community will react badly if we let something die, while promulgating it as useful VAO software. There should of course be a review to decide if it is the time to 'forget' some of these packages, based on matrices, community feedback etc. And, more so, there should be a review to decide and prioritize any eventual improvements (i.e. adding new features). The VAO-SC should have a role in these decisions.